\documentclass[12pt]{iopart}

\usepackage{graphicx}
\usepackage{verbatim}
\usepackage{bm}
\usepackage{lineno}

\newcommand{\beq}{\begin{equation}}
\newcommand{\eeq}{\end{equation}}

\begin{document}

\title[Hebbian plasticity rules abrupt desynchronization in pure simplicial complexes]{Hebbian plasticity rules abrupt desynchronization in pure simplicial complexes}

\author{Ajay Deep Kachhvah, and Sarika Jalan}
\address{Complex Systems Lab, Department of Physics, Indian Institute of Technology Indore - Simrol, Indore - 453552, India}

\vspace{10pt}

\begin{abstract}
This Letter investigates the upshots of adaptive development of pure 2- and 3- simplicial complexes (triad and tetrad) on the nature of the transition to desynchrony of the oscillator ensembles. The adaptation exercised in the pure simplicial coupling takes a cue from the Hebbian learning rule, i.e., the coupling weight of a triad (tetrad) is prone to increase if the oscillators forming it are in phase and decrease if they are out of phase. The coupling weights in these pure simplicial complexes experiencing such adaptation give rise to first-order routes to desynchronization, whose onsets are entirely characterized by respective Hebbian learning parameters. Mean-field analyses presented for the order parameters for the adaptive 2- and 3- simplicial complexes strongly corroborate with the respective numerical assessments.
\end{abstract}

%
\vspace{2pc}
%
%
%
%

\noindent For decades, consideration of the pairwise interaction between different networked dynamical units has been at the forefront to capture the underlying dynamics affecting distinct dynamical processes on various physical and biological complex systems. However, complex systems such as brains~\cite{Petri2014, Benson2018} and social interaction networks~\cite{Iacopini2019, Matamalas2020, Rodriguez2021, Kumar2021} have the underlying topology of higher-order connections. The higher-order interactions are often framed using simplicial complexes~\cite{Salnikov2018, Whitehead1939} that are topological structures. A simplicial complex may include simplices of different dimensions, namely, vertices (0-simplex), edges (1-simplex), triangles (2-simplex), tetrahedrons (3-simplex) and so on stuck to each other.\ A pure simplicial complex is where all facets have the same dimension. Simplicial complexes composed of geometrical objects of different dimensions offer a suitable framework to capture the underlying geometry in complex systems. For instance, simplicial complexes have been used to model the topological map encoded by the hippocampus to capture the environment's basic geometrical features~\cite{Aleksandrov1965,Dabaghian2012}. Recently modeling many-body interactions using simplicial complexes is gaining momentum divulging the rich interplay between network geometry and dynamical processes~\cite{Tanaka2011,Skardal2019,Millan2020,Skardal2020,Lucas2020,Battiston2020,Ghorbanchian2021,Xu2021,Chutani2021,Sun2021,Parastesh2022,Kovalenko2021,Majhi2022,Dai2020}. 

The adaptation is at the backbone of the construction and functioning of many physical and biological complex systems. Experimental findings in neuroscience have led to an impression that the synaptic plasticity between the interacting neurons is the rationale of the learning process and long-term memory in the human brain~\cite{Shimizu2000,Abbott2000}.\ Hebb~\cite{Hebb1949} first conceptualized the fact that the synaptic strength between two interacting neurons is strengthened (weakened) if they are simultaneously firing (not firing), which was later supported by experimental evidence~\cite{Abbott2000,Markram1997,Zhang1998}.\ Further, a neural network can be represented by an ensemble of phase oscillators by encoding the relative spike timing of presynaptic and postsynaptic spikes of the interacting neurons in terms of phases. Thus, the plasticity of the neural network materializes through the adaptation of the synaptic strength between two interacting neurons. The systems of phase oscillators incorporating plasticity in the connection strength between interacting oscillators have revealed intriguing structures and phenomena, for instance, the existence of cluster states~\cite{Seliger2002,Maistrenko207,Niyogi2009,Aoki2009,Chowdhury,Levnajic} or mesoscale structures~\cite{Gutierrez2011,Pitsik2018}, explosive synchronization arising from anti-Hebbian adaptation in monolayer networks~\cite{AGaytan2018} and interlayer Hebbian adaptation in multiplex networks~\cite{Kachhvah2020}.\ Also an adaptive simplicial complex model provides working description of the spatial learning process in the Hippocampal~\cite{Arai2014}. 
\begin{figure}[t!]
	\centering
	\includegraphics[height=4cm,width=8cm]{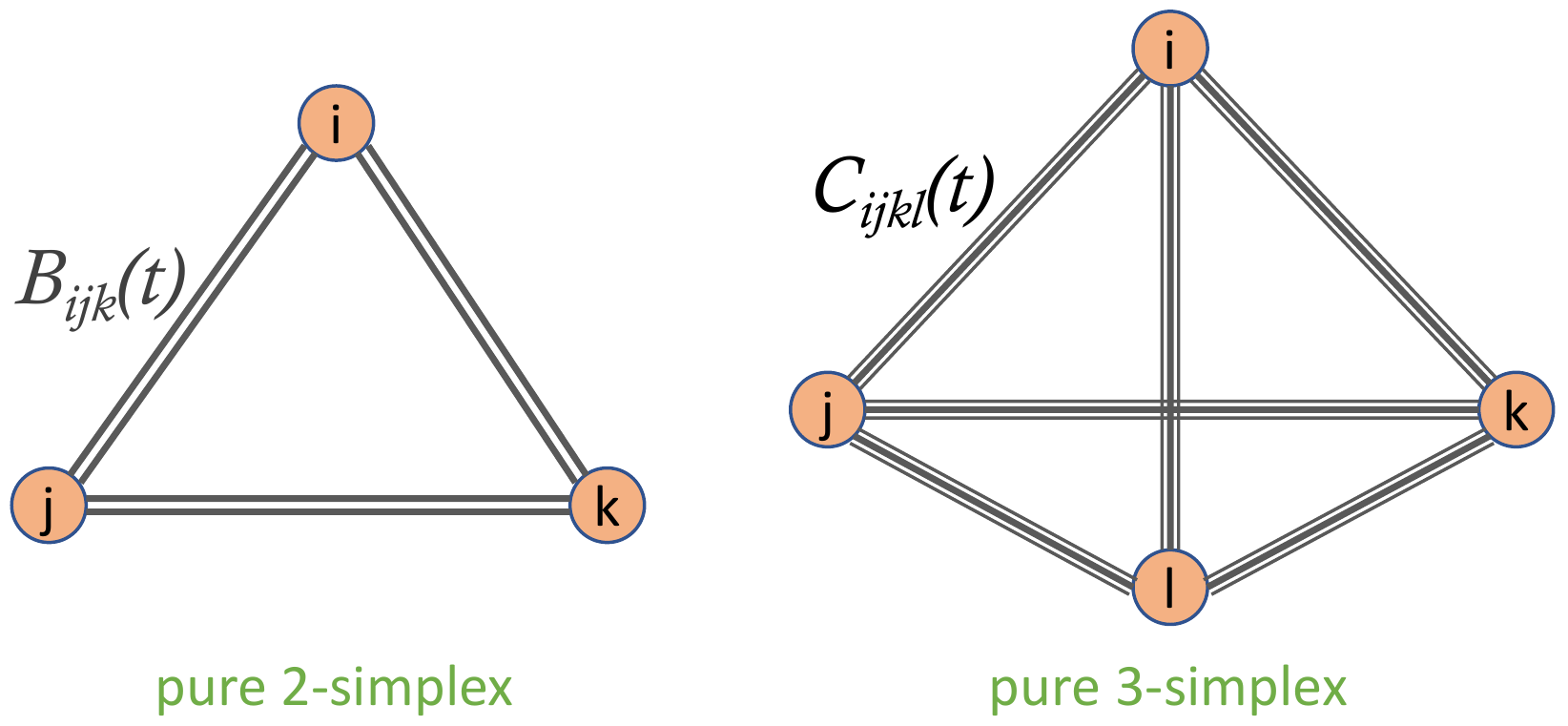}\\
	\vspace{-0.25cm}
	\caption{(Color online) A schematic representation of adaptive pure 2- and 3- simplicial couplings among triad $\{i,j,k\}$ and tetrad $\{i,j,k,l\}$ of nodes, respectively.}
	\label{figure0}
\end{figure}

The collective dynamics of man-made and natural complex systems represented by conventional graphs, for long, have been modeled using pairwise networked Kuramoto oscillators~\cite{Kuramoto1984,Acebron2005}.\ When subjected to distinct adaptation-based features, the pairwise networked Kuramoto oscillators have led to a phenomenon termed first-order or explosive synchronization.\ A first-order transition sports an abrupt jump to the coherence and then an irreversible abrupt collapse to the incoherence as the coupling strength between the interacting oscillators varies~\cite{AGaytan2018,Tanaka1997,Pomerening2003,Gardenes2011,Leyva2012,Danziger2016,Chandrasekar2020,Kuehn2021,Danziger2019,Shepelev2021,Ajay2021,Khanra2021,Frolov}.
In this Letter, breaking away from the traditional approach of assuming adaptation in pairwise coupling, we consider adaptive higher-order coupling interactions among the oscillators. The dynamics of Kuramoto oscillators in the two models studied are respectively driven by pure triadic and pure tetradic couplings that adapt abiding by the Hebbian learning mechanism~\cite{Kachhvah2020}, i.e., a triadic (tetradic) coupling strengthens when the oscillators forming the triad (tetrad) are in phase and weakens when they are out of phase. Such dynamically adaptive pure triadic and tetradic couplings give birth to first-order routes to desynchronization, whose onsets are entirely manageable through respective Hebbian learning parameters.\ Mean-field analyses presented for the 2- and 3- simplicial complexes successfully explain the dependence of the onset points of abrupt desynchronization on the respective Hebbian learning parameters.

To develop an extensive understanding of the repercussions of the Hebbian learning-inspired adaptation in higher-order couplings on synchronization dynamics of oscillator ensembles, we numerically and analytically explore the synchronization dynamics on the pure 2- and 3- simplicial complexes.

\paragraph*{\textbf{Adaptive 2-simplex (triadic) interaction:}}
We consider a phase ensemble of Kuramoto oscillators subjected to adaptive higher-order interactions among them. The higher-order coupling exercised among the oscillators is triadic (three-way) encoded by a 2-simplex (triangle) structure, whose weight dynamically adapt following a mechanism similar to the Hebbian learning rule for pair-wise interaction~\cite{Kachhvah2020}.
The dynamical evolution of a triadic weight assigned to a triplet of nodes $\{i,j,k\}$ having instantaneous phases $\theta_{i,j,k}$ is governed by the Hebbian-learning inspired adaptation
\begin{figure}[t!]
	\centering
	\includegraphics[height=5.5cm,width=8cm]{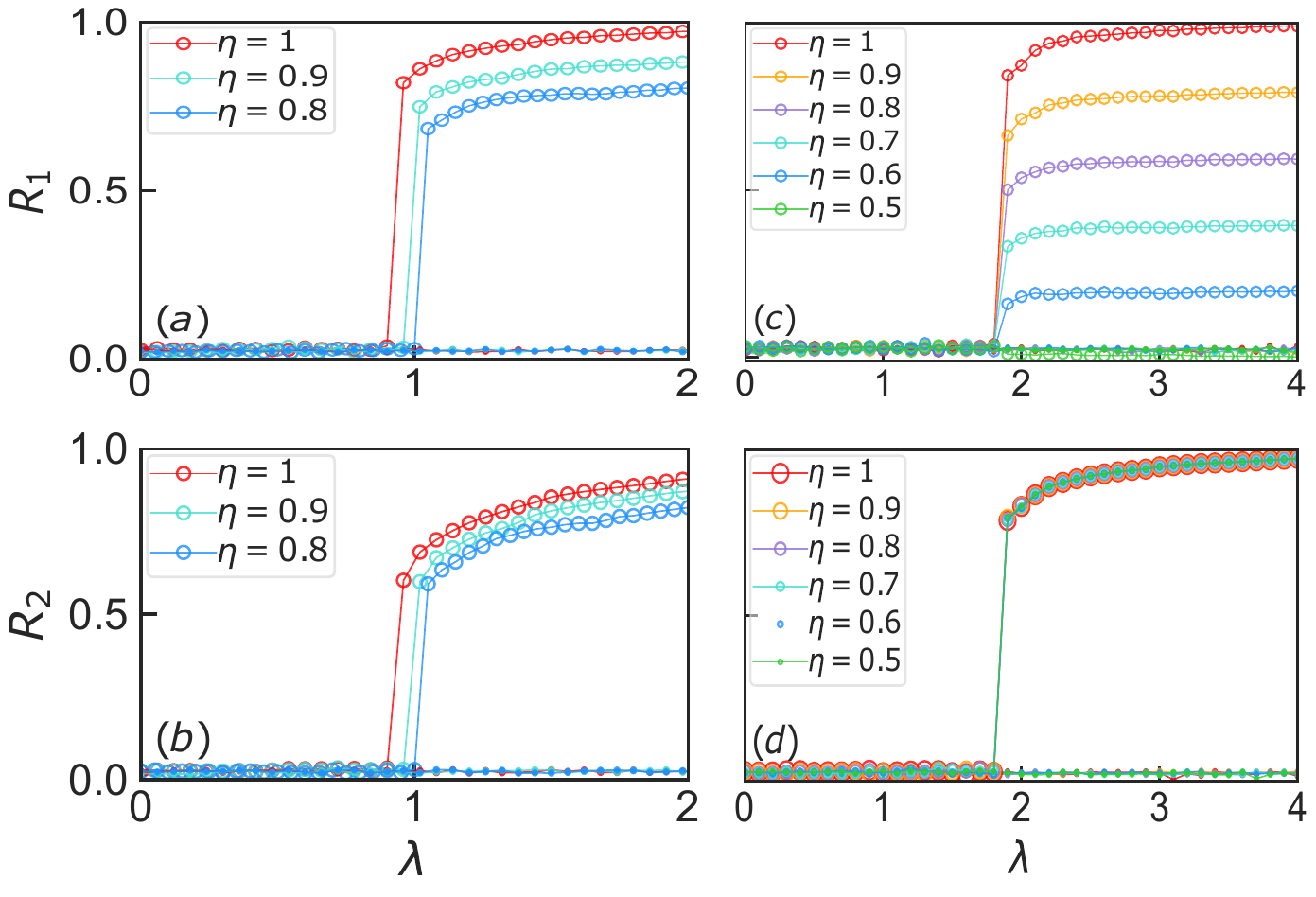}\\
	\vspace{-0.5cm}
	\caption{(Color online) {Robustness of critical coupling against phase asymmetry}: $R_1{-}\lambda$ and $R_2{-}\lambda$ profiles for random 2-simplicial complex of size $N{=}10^3$, $\langle d^{[2]}\rangle{=}12$ when $B_{ijk}$ is static (a-b) and adaptive (c-d) with $\varepsilon{=}1$ and $\mu{=}1$. The Hebbian plasticity negates the impact of initial asymmetry $\eta$ on the outset of desynchronization, instead, makes its impact more profound at the degree of synchronization.}
	\label{figure1}
\end{figure}
\begin{eqnarray}\label{weight_adapt}
\dot B_{ijk}=\varepsilon\cos(\theta_{j}+\theta_{k}-2\theta_{i})-\mu B_{ijk},
\end{eqnarray}
where constants $\varepsilon\in[0,1]$ and $\mu>0$ are learning enhancement rate and saturation rate, respectively. The cosine term implies that the coupling weight $B_{ijk}$ tends to increase if triadic oscillators $\{i,j,k\}$ are in phase, while it tends to decrease if they are out of phase. The linear saturating term $\mu B_{ijk}$ refrains the coupling weight from increasing or decreasing without any bound. The phase evolution of Kuramoto oscillators under the impression of such triadic interaction is ruled by
\begin{eqnarray}\label{phase_evol}
    \dot\theta_{i} = \omega_{i} + \frac{\lambda}{2!\langle d^{[2]}\rangle}\sum\limits_{j,k=1}^{N} B_{ijk} \sin(\theta_{j}+\theta_{k}-2\theta_{i}),
\end{eqnarray}
where natural frequency $\omega_i\ (i{=}1,\dots,N)$ of each oscillator $i$ is drawn randomly from a uniform or unimodal distribution $g(\omega)$, and $\lambda$ is a uniform coupling strength among the interacting nodes. Here, $B$ is an adjacency tensor of 2-simplex that encodes the topology of the network, i.e., $B_{ijk}{=}1$ if there is a triadic coupling among the nodes $\{i,j,k\}$, otherwise $B_{ijk}{=}0$. The number of triangles in the complex a node is part of, is defined as 2-simplex degree $d_i^{[2]} {=} \frac{1}{2!}\sum_{j,k=1}^N B_{ijk}$, and $\langle d^{[2]}\rangle$ denotes the mean 2-simplex degree.

Eqs.~(\ref{weight_adapt}) \& (\ref{phase_evol}) collectively determine the dynamics of the adaptive 2-simplicial complex. To capture the degree of synchronization of the oscillator ensemble, we define generalized order parameter as $R_me^{i\Psi_m} {=}\frac{1}{N}\sum_{j=1}^{N}e^{i m\theta_j}$, where $m{=}1,2$.\ $R_m$ and $\Psi_m$ are amplitude and argument, respectively, corresponding to $m$-cluster order parameter that captures the formation of $m$-clusters.\ $R_1$ quantifies one-cluster synchronization whereas $R_2$ quantifies one-cluster and two-cluster synchronization \cite{Niyogi2009,Karimian2019}. Hence, to measure only two-cluster (anti-phase) synchronization one must adjust $R_2$ for $R_1$, i.e., $|R_2-R_1|$.

Next, we numerically compute the order parameters and discuss the microscopic dynamics of the triadic weights. First, we construct a pure 2-simplex structure by pinpointing each distinct triangle from the 1-simplex (Erd{\"o}s-R{\'e}nyi random graph) structures. The natural frequencies of the nodes are drawn uniform randomly, i.e., $\omega_i{\sim} U(-\Delta, \Delta)$, where $\Delta{=}0.5$ or otherwise mentioned elsewhere.\ For a large $\lambda$, a uniformly selected fraction $\eta$ of the nodes are assigned initial phases to $\theta_i(0){=}0$ and the rest ($1-\eta$) assigned to $\theta_i(0){=}\pi$.\ The initial weights $B_{ijk}(0)$ are then determined by $B_{ijk}(0)=\varepsilon\cos[\theta_{j}(0)+\theta_{k}(0)-2\theta_{i}(0)]$. At first, $\lambda$ is adiabatically decreased until $\lambda{=}0$ and then adiabatically increased. The phase and weight dynamics [Eqs.~(\ref{weight_adapt}) \& (\ref{phase_evol})] are then simulated together on the 2-simplex structures for each $\lambda$ and the order parameters are computed.
\begin{figure}[t!]
	\centering
	\includegraphics[height=3.5cm,width=8cm]{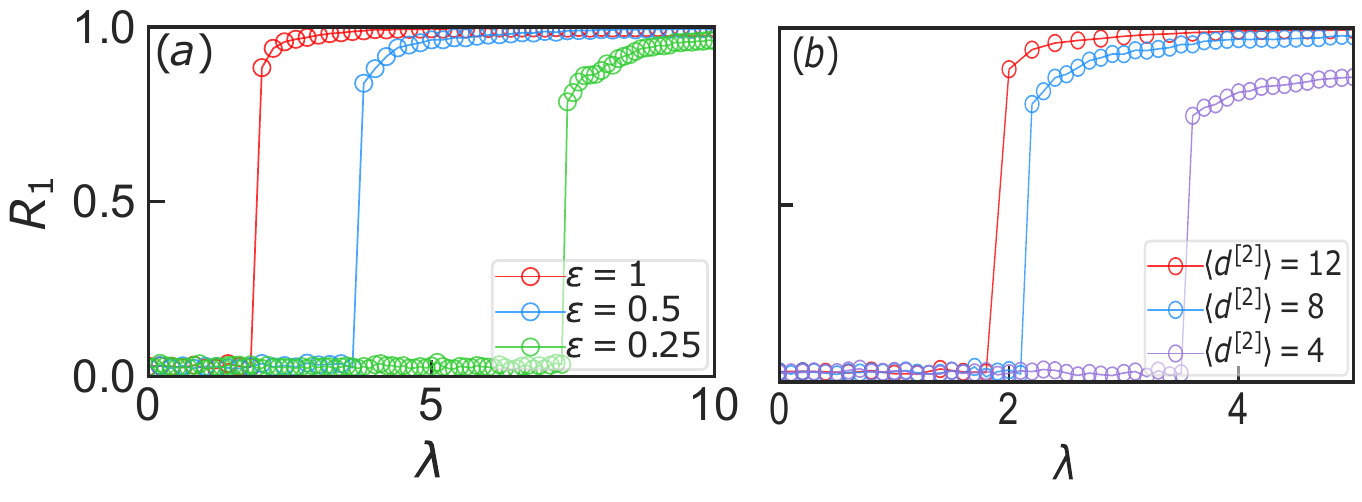}\\
	\vspace{-0.5cm}
	\caption{(Color online) Impact of (a) learning rate $\varepsilon$ for $\langle d^{[2]} \rangle{=}12$ and (b) mean 2-simplex degree $\langle d^{[2]}\rangle$ on transition in a random 2-simplicial complex of $N{=}10^3$, $\mu{=}1$ and $\eta{=}1$. Slower rate $\varepsilon$ requires relatively higher critical $\lambda$ for the system to desynchronize. A higher mean connectivity enhances the level of synchronization and requires relatively lower critical $\lambda$ to desynchronize.}
	\label{figure2}
\end{figure}

\paragraph{The role of Hebbian plasticity:}
The behavior of $R_1$ and $R_2$ as a function of $\lambda$ (see Fig.~\ref{figure1}) illustrates how the onset of abrupt desynchronization is affected under the impression of adaptive evolution of triadic weights. The nature of abrupt desynchronization when the triadic weights $B_{ijk}$ are static [in the absence of Eq.~(\ref{weight_adapt})], is shown in Figs.~\ref{figure1}(a-b). The critical coupling strength $\lambda_c$ at which the abrupt desynchronization takes place for both $R_1$ and $R_2$, increases with the increase in phase asymmetry ($\eta{=}0.5$ yields maximum asymmetry). Under the impression of adaptive triadic weight, both $R_1$ and $R_2$ still adopt the abrupt desynchronization routes, however the critical point $\lambda_c$ is found to remain unaffected of the different values of the fraction $\eta$ as shown in Figs.~\ref{figure1}(c-d). The initial phase asymmetry ($\eta$) does affect the degree of synchronization for $R_1$. The maximum value to which $R_1$ plateaus (at large $\lambda$) decreases with the increase in the level of asymmetry, .i.e., $\eta{=}0.5$ (maximum asymmetry) yields $R_1\simeq0$ while $\eta{=}1$ (symmetry) yields the largest possible $R_1$. However, $R_2$ remains independent of the initial phase asymmetry. Moreover, both $R_1$ and $R_2$ do not synchronize with the forward increase in $\lambda$, hence the incoherent state persists for any $\lambda>0$. These behavior of $R_1$ and $R_2$ also persist for unimodal $g(\omega)$ however at different critical coupling points (see Supplemental Material~\cite{SM}). Note that the Hebbian plasticity of dyadic (1-simplex) interactions does not lead to an abrupt desynchronization (see Supplemental Material~\cite{SM}).
\begin{figure}[t!]
	\centering
	\includegraphics[height=5cm,width=8cm]{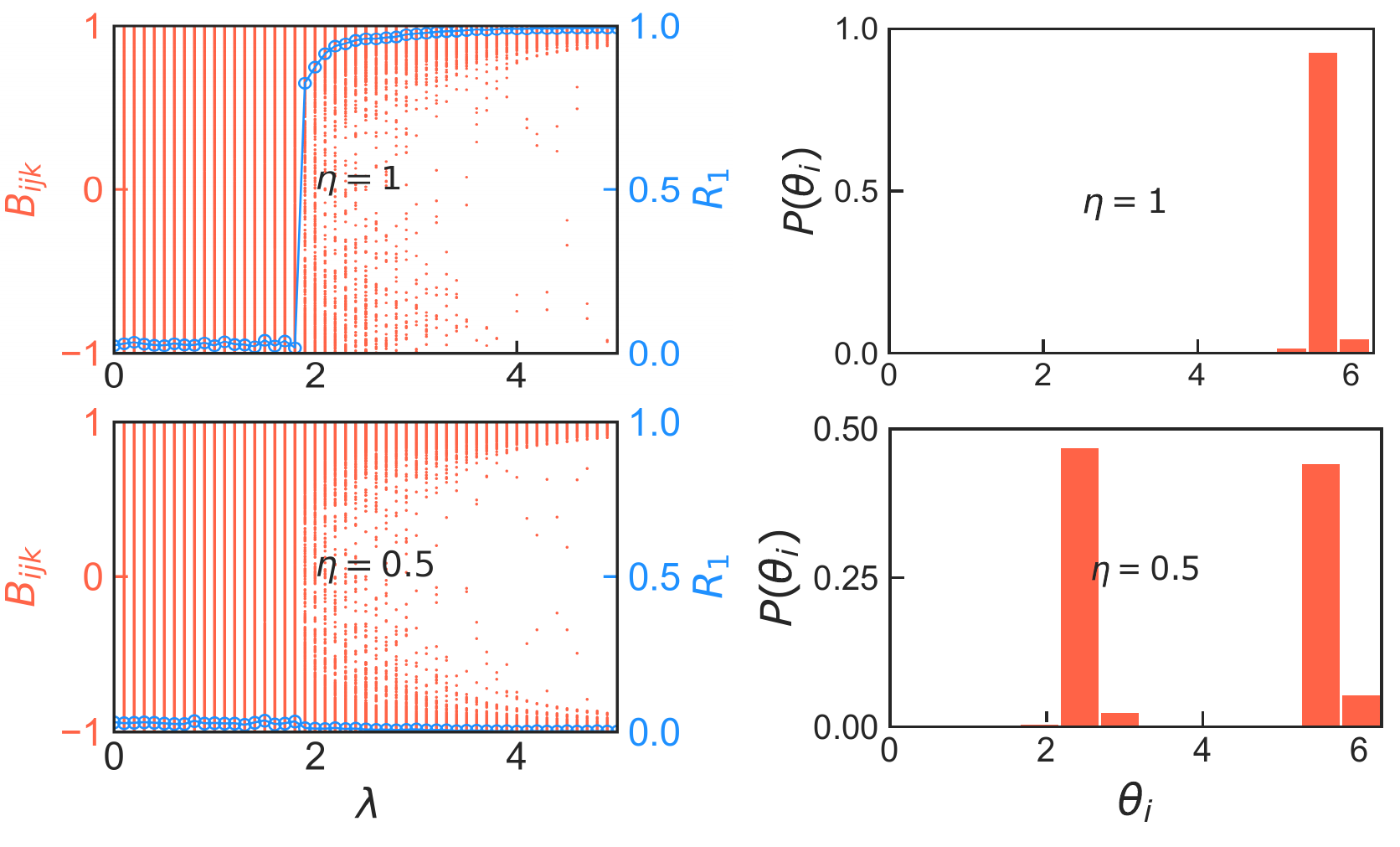}\\
	\vspace{-0.5cm}
	\caption{(Color online) Steady $B_{ijk}$ and steady phases in a random 2-simplicial complex for different values of $\eta$. The distribution $P(\theta_i)$ of steady phases related to $\lambda{=}4$. Other parameters are $N{=}10^3$, $\langle d^{[2]}\rangle{=12}$, $\varepsilon{=}1$ and $\mu{=}1$. The size of two clusters of anti-phases, and that of equal and opposite weights depend on the asymmetry $\eta$.}	
	\label{figure3}
\end{figure}

\paragraph{Impact of learning rate and mean 2-simplex degree:}
Additionally, we demonstrate the effect of learning rate $\varepsilon$ and mean 2-simplex degree $\langle d^{[2]}\rangle$ on the nature of desynchronization in Fig.~\ref{figure2}.\ As the rate $\varepsilon$ is gradually decreased, the onset of first-order desynchronization transition occurs at higher values of critical coupling point $\lambda_c$ [see Fig.~\ref{figure2}(a)]. Thus, the rate $\varepsilon$ provides control over the onset of desynchronization. 
Moreover, the mean triadic connectivity $\langle d^{[2]}\rangle$ also affect the jump height of the abrupt desynchronization [see Fig.~\ref{figure2}(b)].\ As $\langle d^{[2]}\rangle$ is increased, more and more triadic interactions are involved leading to an increment in the height of abrupt jump at rather lower value of critical coupling point.

\paragraph{Stationary triadic weights and phases:}
The microscopic dynamics of stationary weights and stationary phases during the backward sweep of $\lambda$ are presented in Fig.~\ref{figure3} for different values of $\eta$. For $\eta{=}1$, stationary weights are clustered around $B_{ijk}{\rightarrow}\varepsilon/\mu$ at a large $\lambda$. As $\lambda$ is gradually decreased, a gradually increasing number of $B_{ijk}$ starts departing away from $\varepsilon/\mu$ towards $B_{ijk}{=}0$. At the critical point $\lambda_c$, all $B_{ijk}$ get uniformly scattered between $-\varepsilon/\mu$ and $\varepsilon/\mu$ and remain scattered so in the incoherent state until $\lambda{=}0$ is reached. For $\eta{=}1$, the stationary phases corresponding to $\lambda>\lambda_c$ remain locked to form a single peak distribution. For the case of $\eta{=}0.5$, two almost equal-half populations of steady $B_{ijk}$ are flocked around $-\varepsilon/\mu$ and $\varepsilon/\mu$, respectively. A gradual decrease in $\lambda$ shifts a gradually increasing fraction of $B_{ijk}$ towards $0$. At the critical point, the entire $B_{ijk}$ population now gets scattered between $-\varepsilon/\mu$ and $\varepsilon/\mu$. The steady phases related to $\lambda>\lambda_c$ remain almost equally divided into two-clusters at a phase-difference of $\pi$.
\begin{figure}[t]
	\centering
	\includegraphics[height=3cm,width=8cm]{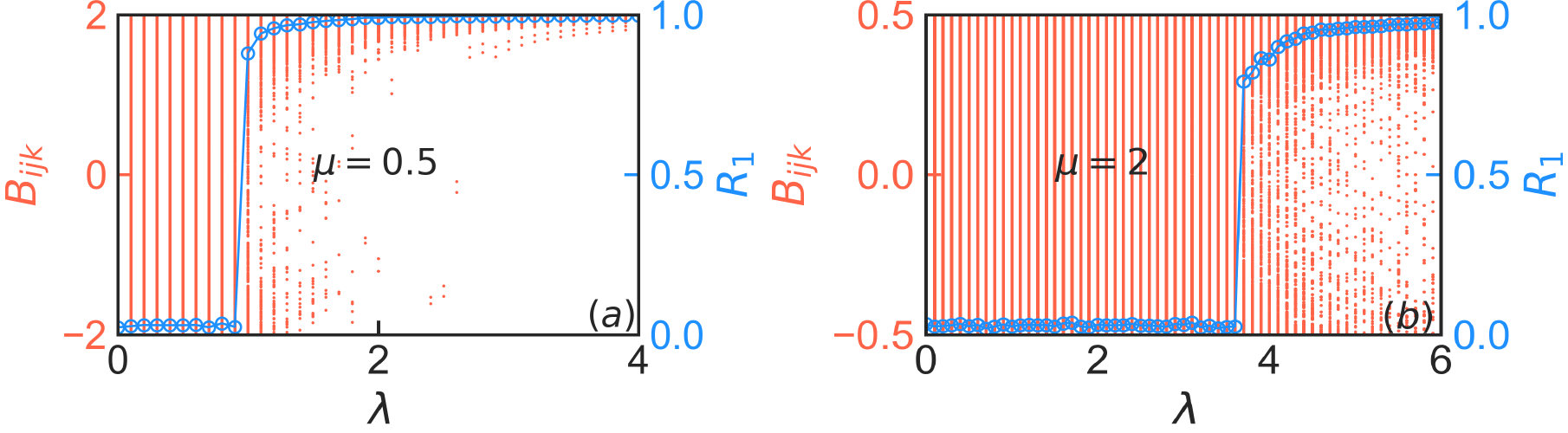}\\
	\vspace{-0.5cm}
	\caption{(Color online) Stationary $B_{ijk}$ for a random 2-simplicial complex for different values of $\mu$. Other parameters are $N{=}10^3$, $\langle d^{[2]}\rangle{=12}$, $\varepsilon{=}1$ and $\eta{=}1$. The upper and lower bounds for the triadic weight $B_{ijk}$ are always $\pm\varepsilon/\mu$, respectively.}
	\label{figure4}
\end{figure}

\paragraph{Impact of saturation rate $\mu$:}
The impact of parameter $\mu$ on stationary triadic weights is shown in Fig.~\ref{figure4}. The ratio ${\varepsilon}/{\mu}$ determines the lower and upper bounds for the stationary weights $B_{ijk}$, which can be corroborated from Eq.~(\ref{weight_adapt}) in the steady state. Parameter $0{<}\mu{\leq}1$ [Fig.~\ref{figure4}(a)] yields higher weight bounds which causes desynchronization at an early coupling strength. On the other hand, $\mu{>}1$ [Fig.~\ref{figure4}(b)] yields lower weight bounds which leads to the desynchronization at rather higher coupling strength. Therefore, $\mu$ also serves as a control parameter in determining the onset of abrupt desynchronization.

\paragraph{Analytical discussion:}
To gain theoretical insight of the underlying dynamics, we turn our focus to an all-to-all connected 2-simplex structure modeled as
\begin{eqnarray}\label{gc_evol}
	\dot\theta_{i} = \omega_{i} + \frac{\lambda}{N^2}\sum_{j,k=1}^{N} B_{ijk} \sin(\theta_{j}+\theta_{k}-2\theta_{i}).
\end{eqnarray}
The system can achieve its stationary state only when both triadic weights and phases achieve their respective stationary states. The stationary triadic weights, $\dot B_{ijk}{=}0$, are given by $B_{ijk}{=}({\varepsilon}/{\mu})\cos(\theta_{j}+\theta_{k}-2\theta_{i})$. Stationary $B_{ijk}$ are set apart into two weight populations, one of $B_{ijk}{\rightarrow}\varepsilon/\mu$ corresponding to $(\theta_{j}+\theta_{k}-2\theta_{i}){=}0$ and other of $B_{ijk}{\rightarrow}-\varepsilon/\mu$ related to $(\theta_{j}+\theta_{k}-2\theta_{i}){=}\pi$. For that matter, an initial asymmetric population comprising $0$ or $\pi$ phases with a probability $\eta$ and $1-\eta$, respectively, is taken into account. The Eq.~(\ref{gc_evol}) can be re-expressed in terms of mean-field parameters by utilizing the expressions for stationary $B_{ijk}$ and $R_2$ as
\begin{equation}
	\dot\theta_{i} = \omega_{i} + bq\sin(2\Psi_2-4\theta_{i});\quad b=\frac{\varepsilon}{2\mu}\quad \&\quad q=\lambda R_2^2.
\end{equation}
\begin{figure}[t!]
	\centering
	\includegraphics[height=5.5cm,width=8cm]{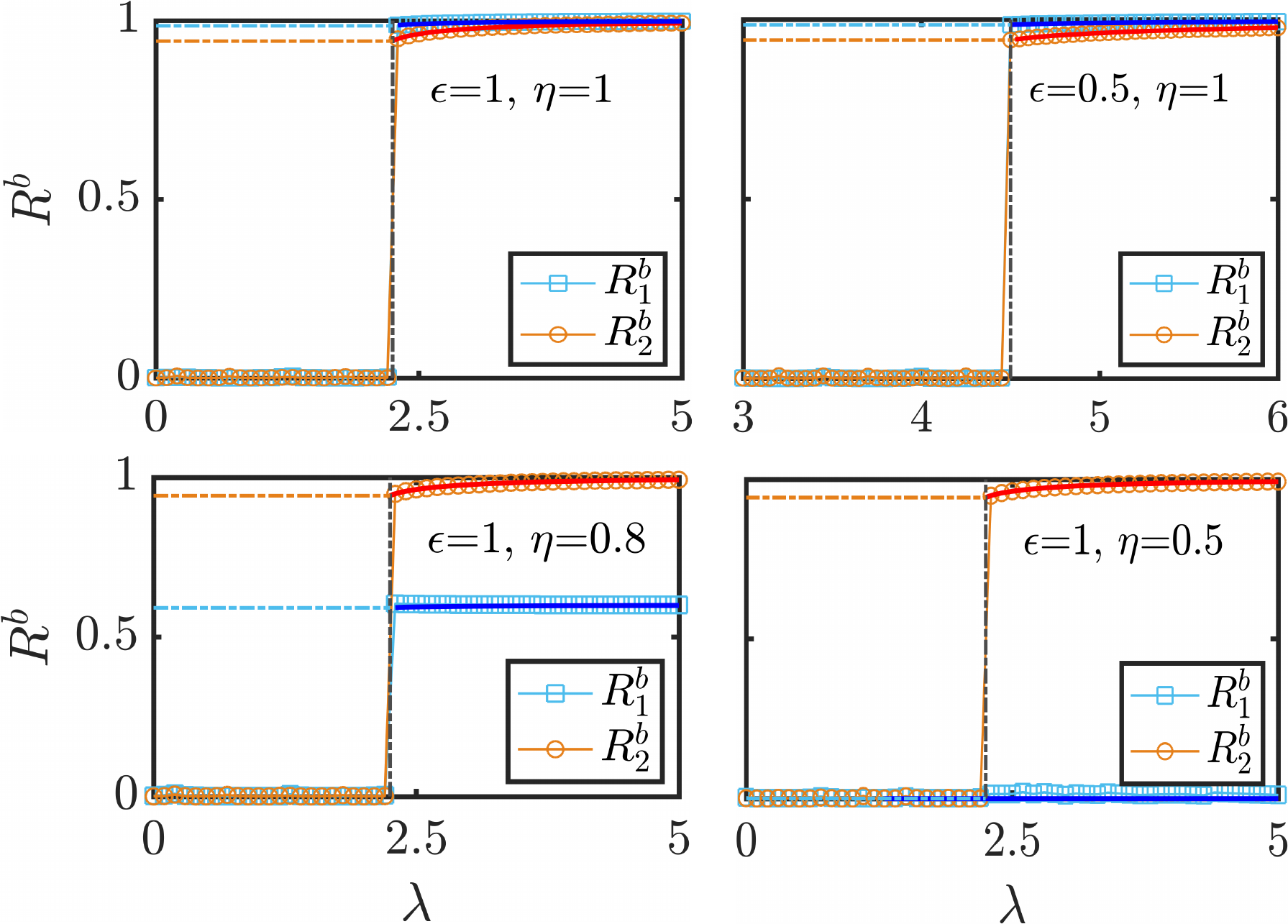}\\
	\vspace{-0.5cm}
	\caption{(Color online) Numerical curves of $R_1{-}\lambda$ and $R_2{-}\lambda$ for an all-to-all connected 2-simplicial complex and respective analytical predictions [Eqs.~(\ref{predict})] in solid blue ($R_1$) and solid red ($R_2$) lines for uniform $\omega_i\sim U(-1,1)$, different values of $\varepsilon$ and $\eta$ while $\mu=1$. The analytical and numerical predictions match quite well.}
	\label{figure5}
\end{figure}
Since phases in a frame rotating with the average intrinsic frequency $\dot\Psi_m$ are given by $\theta_i=\theta_i+\dot\Psi_m t$. Nevertheless, considering a non-rotating frame with $\dot\Psi_m{=}0$ and setting the average phase $\Psi_m{=}0$ by selecting appropriate initial phases, the fixed points in the synchronous state $\dot\theta_i=0$ ($|\omega_i|\leq bq$) are determined by
\begin{eqnarray}\label{fixed_point}
	&\sin(4\theta_i^*) = \frac{\omega_i}{bq},\ \cos(2\theta^*_i)=\pm\sqrt{\frac{1}{2}\Bigg[1+\sqrt{1-\frac{\omega^2_i}{b^2q^2}}\Bigg]},
\end{eqnarray}
where $\eta$ and ($1{-}\eta$) denote a constant probability [$\eta(\omega_i)=\eta$] with which the initial phases are set to $0$ and $\pi$, respectively. Hence, the order parameters for the locked oscillators in continuum limit can be expressed as
\begin{eqnarray}\label{opara_m}
R_m = &\int_{locked}\mathrm{d\omega}g(\omega)\int_{0}^{2\pi}\mathrm{d\theta} e^{\imath m \theta} [\eta\delta(\theta-\theta^\star) + (1{-}\eta) \delta(\theta-\theta^\star-\pi)].
\end{eqnarray}
Further mathematical simplifications supply the real part of the order parameters as
\begin{eqnarray}\label{opara_real}
	&R_1 = (2\eta-1)\int\displaylimits_{|\omega|\leq bq}\mathrm{d\omega}g(\omega) \cos(\theta^*),\nonumber \\
	&R_2 = \int\displaylimits_{|\omega|\leq bq}\mathrm{d\omega}g(\omega) \cos(2\theta^*).
\end{eqnarray}
After plugging in the expressions of $\cos(\theta^*)$ and $\cos(2\theta^*)$ from Eq.~(\ref{fixed_point}) and solving for a uniform distribution $g(\omega)=\frac{1}{2\Delta}$, Eqs.~(\ref{opara_real}) furnish us with (also see Supplemental Material~\cite{SM})
\begin{eqnarray}\label{predict}
	&R_1 = \frac{(2\eta-1)}{2\Delta} \int_{-\Delta}^{\Delta}\mathrm{d\omega} \cos\Bigg[\frac{1}{4}\arcsin\bigg(\frac{\omega}{bq}\bigg)\Bigg],\nonumber \\
	&R_2 = \frac{1}{2\Delta}\int_{-\Delta}^{\Delta}\mathrm{d\omega} \sqrt{\frac{1}{2}\Bigg[1+\sqrt{1-\frac{\omega^2}{b^2q^2}}\Bigg]}.
\end{eqnarray}
It is apparent from Eq.~(\ref{predict}) that only order parameter $R_1$ is impacted by the change in the initial phase asymmetry $\eta$ while $R_2$ remains independent of it. The backward end points of $R_1-\lambda$ and $R_2-\lambda$ traces are obtained by setting $\Delta=bq=b\lambda_c R^2_{2c}$:
\begin{eqnarray}
	R_{1c}=0.98(2\eta-1),\quad R_{2c}=\frac{2\sqrt{2}}{3},\quad \lambda_c=\frac{9\mu\Delta}{4\epsilon}.
\end{eqnarray}
It is apparent that the critical coupling point $\lambda_c$ for the onset of desynchronization is the function of Hebbian plasticity rates $\varepsilon$ and $\mu$.\ The analytical predictions [Eq.~(\ref{predict})] for $R_1$ and $R_2$ are plotted in Fig.~\ref{figure5} for different values of $\varepsilon$ and $\eta$. They are fairly in good agreement with their respective numerical estimations. The values of critical points $\lambda_c$ and $R_{1c}$ and $R_{2c}$ are also shown matching their respective numerical assessments.

\paragraph*{\textbf{Adaptive 3-simplex (tetradic) interaction:}}
Now we consider the higher-order coupling interaction among the oscillators to be tetradic encoded by a pure $3$-simplex (tetrahedron) structure. The weights of the tetradic coupling also dynamically adapt in a fashion similar to the Hebbian learning rule. The evolution of the phases and the tetradic weights can respectively be expressed as
\begin{figure}[t!]
	\centering
	\includegraphics[height=5.5cm,width=8cm]{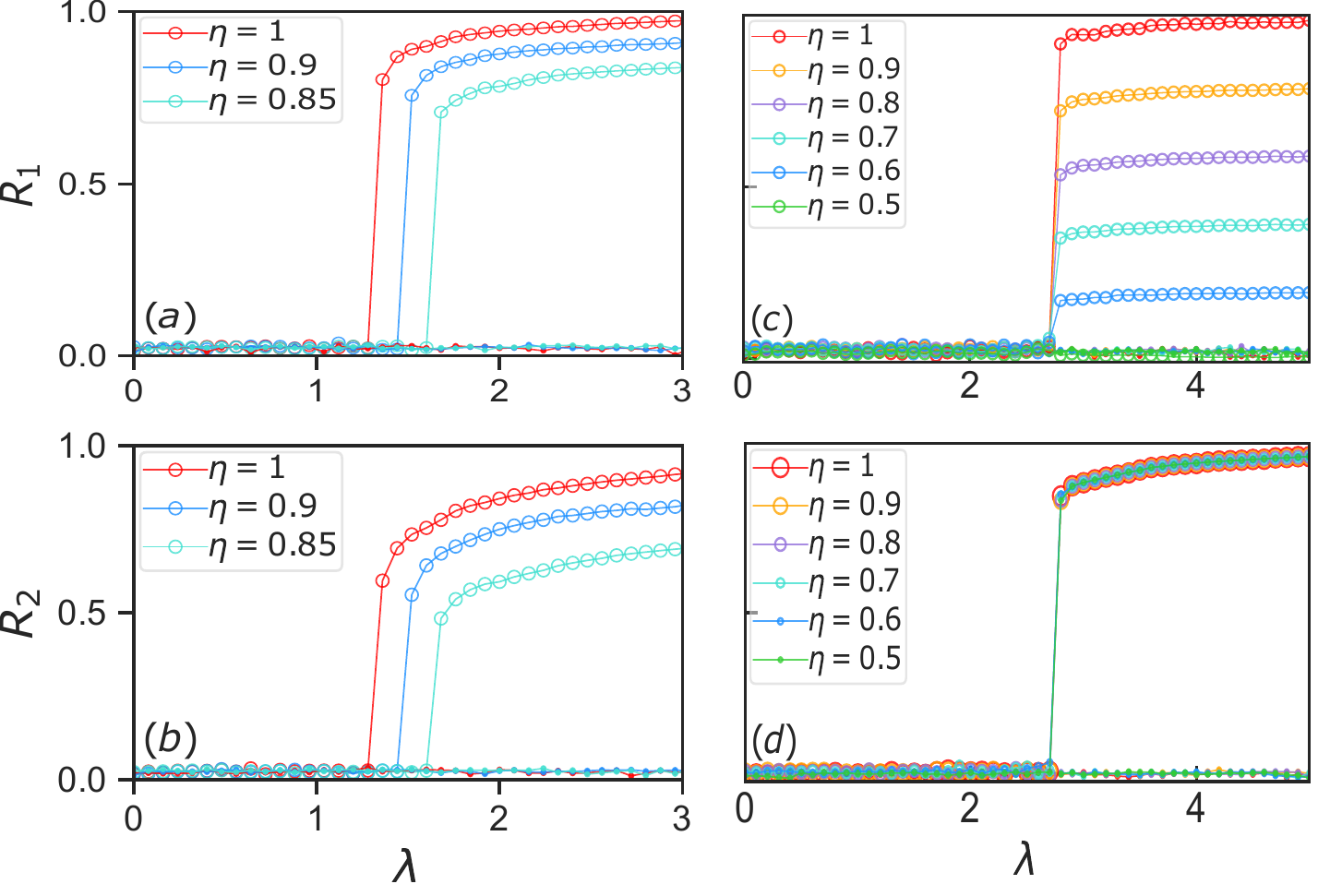}\\
	\vspace{-0.5cm}
	\caption{(Color online) {Robustness of critical coupling against asymmetry}: $R_1{-}\lambda$ and $R_2{-}\lambda$ profiles for random 3-simplices of size $N{=}10^3$, $\langle d^{[3]}\rangle{=}12$ and uniform $\omega_i\sim U(-0.5,0.5)$ when $C_{ijkl}$ are static (a-b) and adaptive (c-d) with $\sigma{=}1$ and $\nu{=}1$. The Hebbian plasticity makes the impact of $\eta$ more profound at the level of synchronization but negates its on the outset of desynchronization.}
	\label{figure6}
\end{figure}
\begin{eqnarray}\label{tetradHebb}
&\dot\theta_{i} = \omega_{i} + \frac{\lambda}{3!\langle d^{[3]}\rangle}\sum_{j,k,l=1}^{N}C_{ijkl} \sin(\theta_j+\theta_k+\theta_l-3\theta_{i}), \nonumber\\
&\dot C_{ijkl} = \sigma\cos(\theta_j+\theta_k+\theta_l-3\theta_{i}) - \nu C_{ijkl},
\end{eqnarray}
where $C$ is 3-simplex adjacency tensor encoding a network of tetrahedrons. $C_{ijkl}{=}1$ if a tetrahedronic connection exists among the nodes $\{i, j, k, l\}$, otherwise $C_{ijkl}{=}0$. 3-simplex degree of a node $d_i^{[3]} {=} \frac{1}{3!}\sum_{j,k,l=1}^N C_{ijkl}$ denotes the number of tetrahedrons it is part of.\ We consider the 1-simplex structure (Erd{\"o}s-R{\'e}nyi random graph) and then identify each distinct tetrahedron to construct a pure 3-simplicial complex.

The evolution of the system begins at a large $\lambda$ with initial phases fixed to $0$ and $\pi$ with a constant probability $\eta$ and $1-\eta$, respectively. The initial tetradic weights are assigned as $C_{ijkl}(0){=}\sigma\cos[\theta_{j}(0){+}\theta_{k}(0){+}\theta_{l}(0){-}3\theta_{i}(0)]$. The nature of abrupt desynchronization when the tetradic weights $C_{ijkl}$ are static [see Figs.~\ref{figure6}(a-b)] and adaptive [see Figs.~\ref{figure6}(c-d)]. For static $C_{ijkl}$ (${=}1$), the decrease in $\eta$ advances the onset point $\lambda_c$ for both backward $R_1$ and $R_2$ transitions. However under the impact of adaptive $C_{ijkl}$, both $R_1$ and $R_2$ continue to espouse backward abrupt transition routes but at a fixed $\lambda_c$ for different values of $\eta$. $R_2$ remains independent of $\eta$ whereas $R_1$ decreases with decrease in the value of $\eta$. Also, both $R_1$ and $R_2$ always exhibit incoherence for forward continuation in $\lambda$ be it either static or adaptive weight $C_{ijkl}$.

\begin{figure}[t!]
	\centering
	\includegraphics[height=3.5cm,width=8cm]{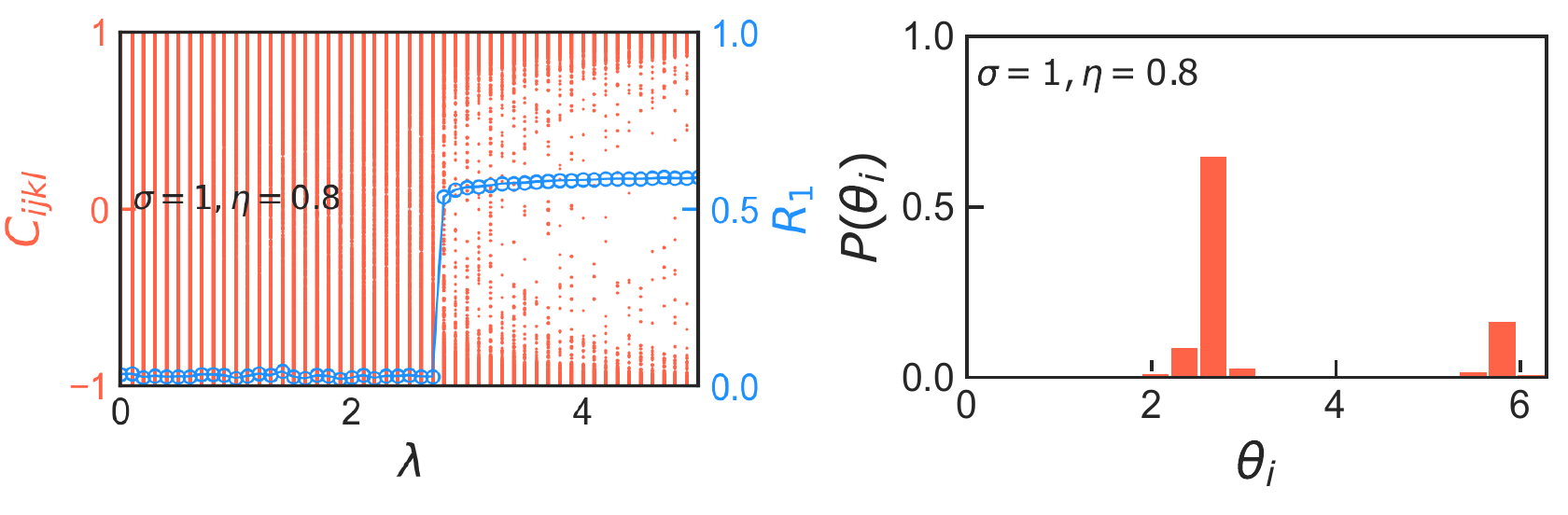}\\
	\vspace{-0.5cm}
	\caption{(Color online) Stationary $C_{ijkl}$ and related stationary phases at $\lambda=4.5$ for a 3-simplicial complex for uniform $\omega_i\sim U(-0.5,0.5)$, $N{=}10^3$, $\langle d^{[3]}\rangle{=}12$, $\sigma{=}1$, $\nu{=}1$ and $\eta={0.8}$. The size of two clusters of anti-phases (equal and opposite weights) depends on the asymmetry $\eta$.}
	\label{figure7}
\end{figure}

Now we look at the nature of stationary weights $\dot C_{ijkl}{=}0$, $C_{ijkl} = (\sigma/\nu)\cos(\theta_j+\theta_k+\theta_l-3\theta_{i})$. Fig.~\ref{figure7} unveils that all $C_{ijkl}$ in the synchronous state are segregated into two clusters, one of $C_{ijkl}{\rightarrow}\sigma/\nu$ related to $(\theta_j+\theta_k+\theta_l-3\theta_{i}){\rightarrow}0$ and other of $C_{ijkl}{\rightarrow}-\sigma/\nu$ related to $(\theta_j+\theta_k+\theta_l-3\theta_{i}){\rightarrow}\pi$. The size of two anti-phase clusters depends on parameter $\eta$. In the asynchronous state, all $C_{ijkl}$ are uniformly scattered between $-\sigma/\nu$ and $\sigma/\nu$. 

For the sake of analytical simplification we consider an all-to-all connected pure 3-simplicial complex, for which the interaction term in the phase evolution [Eq.~(\ref{tetradHebb})] is normalized by $N^3$ instead of $6\langle d^{[3]}\rangle$.\ Further, substituting $C_{ijkl}$ with the stationary $C_{ijkl}$, the phase evolution then can be rewritten in terms of $R_2$ and $\Psi_2$.
\begin{eqnarray}
	\dot\theta_i=\omega_i+d\lambda R_2^3\sin(3\Psi_2-6\theta_i);\quad d=\frac{\sigma}{2\nu}.
\end{eqnarray}
Consider a non-rotating frame with mean frequency $\dot\Psi_2{=}0$ and mean phase $\Psi_2{=}0$, the fixed points of the locked oscillators are then supplied by $\dot\theta_i{=}0$, i.e., $\sin(6\theta_i^*){=}\omega_i/d\lambda R_2^3$.
Eq.~(\ref{opara_m}) when worked out for a viable fixed point and a uniform distribution $g(\omega){=}\frac{1}{2\Delta}$, furnishes us with the following order parameters (see Supplemental Material~\cite{SM})
\begin{figure}[t!]
	\centering
	\includegraphics[height=5.5cm,width=8cm]{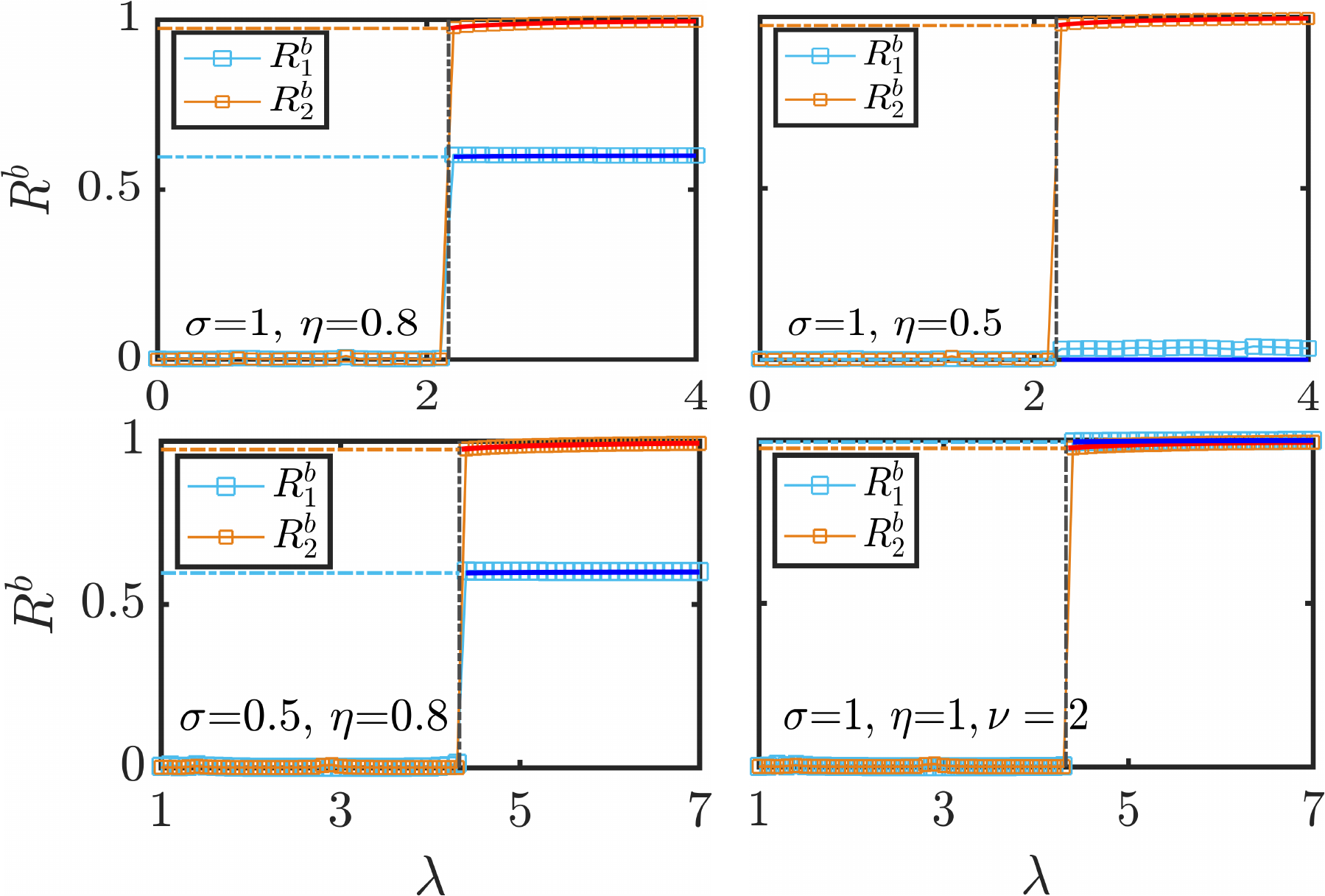}\\
	\vspace{-0.5cm}
	\caption{(Color online) Numerical branches for $R_1{-}\lambda$ and $R_2{-}\lambda$ for the all-to-all connected 3-simplicial complex and respective analytical predictions [Eqs.(\ref{tetrad_R1R2})] in solid blue ($R_1$) and solid red ($R_2$) lines for uniform $\omega_i\sim U(-1,1)$, different values of $\sigma$ and $\eta$ while $\nu=1$. The analytical predictions and numerical results are in good agreement.}
	\label{figure8}
\end{figure}
\begin{eqnarray}\label{tetrad_R1R2}
	&R_1 =\frac{(2\eta-1)}{2\sqrt2 \Delta}\int_{-\Delta}^{\Delta} d\omega \sqrt{1+\frac{1}{2}\bigg[\sqrt[3]{x+iy} + \sqrt[-3]{x+iy}\bigg]},\nonumber\\
	&R_2=\frac{1}{2\Delta}\int_{-\Delta}^{\Delta} d\omega \frac{1}{2}\bigg[\sqrt[3]{x+iy} + \sqrt[-3]{x+iy}\bigg], 
\end{eqnarray}
where $x{=}\sqrt{1-{\omega^2}/{u^2}},\ y{=}{\omega}/{u}$.
Eqs.~(\ref{tetrad_R1R2}) for $\Delta=u=d\lambda R_2^3$ supply us with the backward critical points:
\begin{eqnarray}
	&R_{2c}=\frac{9\sqrt{3}}{16},\quad R_{1c}=0.99 (2\eta-1),\quad \lambda_c=2.16\frac{\Delta\nu}{\sigma}.
\end{eqnarray}
Hence the critical coupling points for the outset of desynchronization is controlled by the Hebbian learning parameters $\sigma$ and $\nu$.\ The theoretical predictions for $R_1$ and $R_2$ [Eq.~(\ref{tetrad_R1R2})] and their numerical estimations fall in fine concurrence with each other for different values of $\sigma$ and $\eta$ (see Fig.~\ref{figure8}), and so do the hypothesized critical points $R_{1c}$, $R_{2c}$ and $\lambda_c$.

\paragraph*{Conclusion:}
We studied desynchronization transition in oscillator ensembles networked in triadic (pure 2-simplex) and  tetradic (pure 3-simplex) structures with adaptive coupling. The adaptation exercised in triadic (tetradic) coupling is inspired by the Hebbian learning mechanism, i.e., a triadic (tetradic) weight tends to increase when the nodes forming the triad (tetrad) are in phase and decrease when they are out of phase. The dynamical evolution of such Hebbian learned triadic (tetradic) coupling leads to a first-order desynchronization in the network. For both pure 2- and 3- simplicial complexes, respective triggering points of the emergent abrupt desynchronization are manageable through respective Hebbian learning parameters, i.e., the learning rate and saturation rate. Besides, for both the complexes, the initial phase asymmetry characterizes the one-cluster order parameter while the two-clusters order parameter remains independent of it. Also, the triadic (tetradic) weights remain uniformly distributed in the incoherent state, whereas they form one-cluster or two-clusters depending on the initial phase asymmetry. The mean-field analyses presented for pure 2- and 3- simplicial complexes too successfully explain the respective Hebbian adaptation governed nature of one-cluster and two-clusters order parameters.

So far, all the investigations featuring the Hebbian learning mechanism have revolved around employing a pair-wise interaction between the oscillators. The present study would be advantageous in better understanding the role of adaptive higher-order neural synaptic connections in the learning process and procuring long term memory in the brain. The proposed models also open an avenue for investigating the imprint of concurrent adaptation in various combinations of dyadic, triadic and tetradic couplings on the synchronization dynamics of oscillator ensembles.

\ack
SJ acknowledges DST POWER grant SERB/F/9035/2021-2022.

\section*{References}

\begin{thebibliography}{99}

	\bibitem{Petri2014} G. Petri, P. Expert, F. Turkheimer, R. Carhart-Harris, D. Nutt, P. J. Hellyer and F. Vaccarino 2014 {\em J. R. Soc. Interface} {\bf 11} 20140873

	\bibitem{Benson2018} Benson A R, Abebe R, Schaub M T, Jadbabaie A and Kleinberg J  2018 {\em Proc. Natl. Acad. Sci. U.S.A.} {\bf 115} E11221 
	
	\bibitem{Iacopini2019} I. Iacopini, G. Petri, A. Barrat, and V. Latora 2019 {\em Nat. Commun.} {\bf 10} 2485 
	
	\bibitem{Matamalas2020}  J. T. Matamalas, S. Gómez, and A. Arenas 2020 {\em Phys. Rev. Research} {\bf 2} 012049(R) 
	
	\bibitem{Rodriguez2021} U. Alvarez-Rodriguez, F. Battiston, G. Ferraz de Arruda, Y. Moreno, M. Perc, and V. Latora 2021 {\em Nat. Hum. Behav.} {\bf 5} 586 
 	
 	\bibitem{Kumar2021} A. Kumar, S. Chowdhary, V. Capraro, and M. Perc 2021 {\em Phys. Rev. E} {\bf 104} 054308 

	\bibitem{Salnikov2018} V. Salnikov, D. Cassese, and R. Lambiotte 2018 {\em Eur. J. Phys.} {\bf 40} 014001 
	
	\bibitem{Whitehead1939} J. H. C. Whitehead 1939 {\em Proc. London Math. Soc.} {\bf 45} 243 
	
	\bibitem{Aleksandrov1965} P. S. Aleksandrov 1961 {\em Elementary concepts of topology} (New York: Dover)
	
	\bibitem{Dabaghian2012} Y. Dabaghian, F. M{\'e}moli, L. Frank, and G. Carlsson 2012 {\em PLOS Comput. Biol.} {\bf 8} e1002581 
	
	\bibitem{Tanaka2011} T. Tanaka, and T. Aoyagi 2011 {\em Phys. Rev. Lett.} {\bf 106} 224101 
	
	\bibitem{Skardal2019} P. S. Skardal, and A. Arenas 2019 {\em Phys. Rev. Lett.} {\bf 122} 248301 
	
	\bibitem{Millan2020} A. P. Mill{\'a}n, J. J. Torres, and G. Bianconi 2020 {\em Phys. Rev. Lett.} {\bf 124} 218301 
	
	\bibitem{Skardal2020} P. S. Skardal, and A. Arenas 2020 {\em Commun. Phys.} {\bf 3} 218 
	
	\bibitem{Lucas2020} M. Lucas, G. Cencetti, and F. Battiston 2020 {\em Phys. Rev. Research} {\bf 2} 033410 
	
	\bibitem{Battiston2020} F. Battiston, G. Cencetti, I. Iacopini, V. Latora, M. Lucas, A. Patania, Jean-Gabriel Young, and G. Petri 2020 {\em Phys. Rep.} {\bf 874} 1 
	
	\bibitem{Ghorbanchian2021} R. Ghorbanchian, J. G. Restrepo, J. J. Torres, and G. Bianconi 2021 {\em Commun. Phys.} {\bf 4} 120 
	
	\bibitem{Xu2021} C. Xu, X. Tang, H. L{\"u}, K. Alfaro-Bittner, S. Boccaletti, M. Perc, and S. Guan 2021 {\em Phys. Rev. Research} {\bf 3} 043004; C. Xu, and P. S. Skardal 2021 {\em Phys. Rev. Research} {\bf 3} 013013; C. Xu, X. Wang, and P. S. Skardal 2020 {\em Phys. Rev. Research} {\bf 2} 023281 
	
	\bibitem{Chutani2021} M. Chutani, B. Tadi{\'c}, and N Gupte 2021 {\em Phys. Rev. E} {\bf104} 034206 
	
	\bibitem{Sun2021} H. Sun, and G. Bianconi 2021 {\em Phys. Rev. E} {\bf 104} 034306
	
	\bibitem{Parastesh2022} F. Parastesh, M. Mehrabbeik, K. Rajagopal, Sajad Jafari, and M. Perc 2022 {\em Chaos} {\bf 32} 013125 
	
	\bibitem{Kovalenko2021} K. Kovalenko, X. Dai, K. Alfaro-Bittner, A. M. Raigorodskii, M. Perc, and S. Boccaletti 2021 {\em Phys. Rev. Lett.} {\bf 127} 258301 
	
	\bibitem{Majhi2022} S. Majhi , M. Perc, and D. Ghosh 2022 {\em J. R. Soc. Interface} {\bf 19} 20220043 
	
	\bibitem{Dai2020} X. Dai, X. Li, H. Guo, D. Jia, M. Perc, P. Manshour, Z. Wang, and S. Boccaletti 2020 {\em Phys. Rev. Lett.} {\bf 125} 194101 
			
	\bibitem{Shimizu2000} E. Shimizu, Y. P. Tang, C. Rampon, and J. Z. Tsien 2000 {\em Science} {\bf 290} 1170 
	
	\bibitem{Abbott2000} L. F. Abbott, and S. B. Nelson 2000 {\em Nat. Neurosci.} {\bf 3} 1178 
	
	\bibitem{Hebb1949} D. O. Hebb 1949 {\em The Organization of Behavior} (New York: John Wiley \& Sons) 
	
	\bibitem{Markram1997} H. Markram, J. L{\"u}bke, M. Frotscher, and B. Sakmann 1997 {\em Science} {\bf 275} 213 
	
	\bibitem{Zhang1998} L. I. Zhang, H. W. Tao, C. E. Holt, W. A. Harris, and M. Poo 1998 {\em Nature} {\bf 395} 37 
	
	\bibitem{Seliger2002} P. Seliger, S. C. Young, and L. S. Tsimring 2002 {\em Phys. Rev. E} {\bf 65} 041906
	
	\bibitem{Maistrenko207} Y. L. Maistrenko, B. Lysyansky, C. Hauptmann, O. Burylko, and P. A. Tass 2007 {\em Phys. Rev. E} {\bf 75} 066207 
	
	\bibitem{Niyogi2009} R. K. Niyogi, and L. Q. English 2009 {\em Phys. Rev. E} {\bf 80} 066213 
	
	\bibitem{Aoki2009} T. Aoki, and T. Aoyagi 2009 {\em Phys. Rev. Lett.} {\bf 102} 034101 
 	
 	\bibitem{Chowdhury} S. N. Chowdhury, S. Rakshit, J. M. Buldú, D. Ghosh, and C. Hens 2021 {\em Phys. Rev. E} {\bf 103} 032310; S. N. Chowdhury, D. Ghosh, and C. Hens 2020 {\em Phys. Rev. E} {\bf 101} 022310
 	
 	\bibitem{Levnajic} Z. Levnaji{\'c} 2012 {\em Sci. Rep.} {\bf 2} 967; Z. Levnaji{\'c} 2011 {\em Phys. Rev. E} {\bf 84} 016231
 	
 	\bibitem{Gutierrez2011} R. Guti\'errez, A. Amann, S. Assenza, J. G\'omez-Garde\~nes, V. Latora, and S. Boccaletti 2011 {\em Phys. Rev. Lett.} {\bf 107} 234103 
	
	\bibitem{Pitsik2018} E. Pitsik, V. Makarov, D. Kirsanov, N. Frolov, M. Goremyko, X. Li, Z. Wang, A. Hramov, and S. Boccaletti 2018 {\em New J. Phys.} {\bf 20} 075004 
	
	\bibitem{AGaytan2018} V. Avalos-Gayt\'an, J. A. Almendral, I. Leyva, F. Battiston, V. Nicosia, V. Latora, and S. Boccaletti 2018 {\em Phys. Rev. E} {\bf 97} 042301 
  
	\bibitem{Kachhvah2020} A. D. Kachhvah, X. Dai, S. Boccaletti, and S Jalan 2020 {\em New J. Phys.} {\bf 22} 122001 
	
	\bibitem{Arai2014} M. Arai, V. Brandt, and Y. Dabaghian 2014 {\em PLOS Comput. Biol.} {\bf 10} e1003651 
	
	\bibitem{Kuramoto1984} Y. Kuramoto 1975 {\em International Symposium on Mathematical Problems in Theoretical Physics, Lecture Notes in Physics} {\bf 39} 

	\bibitem{Acebron2005} J. A. Acebr\'on, L. L. Bonilla, C. J. P\'erez Vicente, F. Ritort, and R. Spigler 2005 {\em Rev. Mod. Phys.} {\bf 77} 137 
  
	\bibitem{Tanaka1997} H. A. Tanaka, A. J. Lichtenberg, and S. Oishi 1997 {\em Phys. Rev. Lett.} {\bf 78} 2104 

	\bibitem{Pomerening2003} J. R. Pomerening, E. D. Sontag, and Jr J. E. Ferell 2003 {\em Nat. Cell Biol.} {\bf 5} 346 
	
	\bibitem{Gardenes2011} J. G\'omez-Garde\~nes, S. G\'omez, A. Arenas, and Y. Moreno 2011 {\em Phys. Rev. Lett.} {\bf 106} 128701 

	\bibitem{Leyva2012} I. Leyva, R. Sevilla-Escoboza, J. M. Buld\'u, N. I. Sendi\~na, J. G\'omez-Garde\~nes, A. Arenas, Y. Moreno, S. G\'omez, R. Jaimes-Re\'ategui, and S. Boccaletti 2012 {\em Phys. Rev. Lett.} {\bf 108} 168702 

	\bibitem{Danziger2016} M. M. Danziger, O. I. Moskalenko, S. A. Kurkin, X. Zhang, S. Havlin, and S. Boccaletti 2016 {\em Chaos} {\bf 26} 065307 

	\bibitem{Chandrasekar2020} V. K. Chandrasekar, M. Manoranjani, and S. Gupta 2020 {\em Phys. Rev. E} {\bf 102} 012206 
  
	\bibitem{Kuehn2021} C. Kuehn, and C. Bick 2021 {\em Sci. Adv.} {\bf 7} eabe3824 

	\bibitem{Danziger2019} M. M. Danziger, I. Bonamassa, S. Boccaletti, and S. Havlin 2019 {\em Nat. Phys.} {\bf 15} 178 

	\bibitem{Shepelev2021} I. Shepelev, A. Bukh, G. Strelkova, and V. Anishchenko 2021 {\em Chaos Soliton Fract.} {\bf 143} 110545 

	\bibitem{Ajay2021} A. D. Kachhvah, and S. Jalan 2021 {\em Phys. Rev. E} {\bf 104} L042301 

	\bibitem{Khanra2021} P. Khanra, and P. Pal 2021 {\em Chaos Soliton Fract.} {\bf 143} 110621 

	\bibitem{Frolov} N. Frolov, and A. Hramov 2021 {\em Chaos} {\bf 31}, 063103; N. Frolov, and A. Hramov 2021 {\em Chaos Soliton Fract.} {\bf 147} 110955 
	
	\bibitem{Karimian2019} M. Karimian, D. Dibenedetto, M. Moerel, T. Burwick, R. L. Westra,  P. De~Weerd, and M. Senden 2019 {\em Chaos} {\bf 29} 083122 
	
	\bibitem{SM} See Supplemental Material concerning synchronization profiles for the Hebbian plasticity adapted dyadic interaction, abrupt desynchronization for unimodal frequency distributions, and the steps deriving exact expressions for the order parameters related to the adaptive 2- and 3- simplicial complexes. 
		
\end{thebibliography}
\providecommand{\newblock}{}

\end{document}


\title{Supplemental Material: Hebbian plasticity rules abrupt desynchronization in pure simplicial complexes}

\author{Ajay Deep Kachhvah, and Sarika Jalan}
\affiliation{Complex Systems Lab, Department of Physics, Indian Institute of Technology Indore, Khandwa Road, Simrol, Indore-453552, India}

\maketitle

This Supplemental Material includes auxiliary figures and mathematical steps, which support the results and arguments described in the main text.

\subsection{Hebbian plasticity of dyadic interaction}
The phase and weight evolutions of an all-to-all connected 1-simplex (dyadic) network of $N$ Kuramoto oscillators, with its couplings subjected to the Hebbian plasticity (learning rate $\alpha$ and saturation rate $\kappa$), are given by
\begin{eqnarray}\label{gc_evol}
	\dot\theta_{i} = \omega_{i} + \frac{\lambda}{N}\sum_{j=1}^{N} A_{ij} \sin(\theta_{j}-\theta_{i}),\\
\dot A_{ij}=\alpha\cos(\theta_{j}-\theta_{i})-\kappa A_{ij},
\end{eqnarray}
respectively.
At a large $\lambda$, the oscillators are uniform randomly assigned phases $\theta_i(0){=}0$ and $\pi$ with probability $\eta$ and $1-\eta$, respectively. The initial coupling weights are then determined by $A_{ij}(0)=\alpha\cos[\theta_{j}(0)-\theta_{i}(0)]$.
\begin{figure}[ht!]
	\centering
	\begin{tabular}{cc}
	\includegraphics[height=3.5cm,width=5cm]{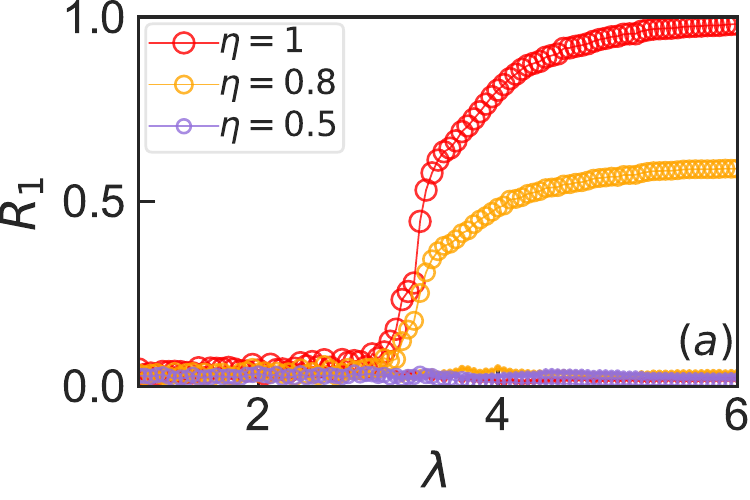}&
	\includegraphics[height=3.5cm,width=5cm]{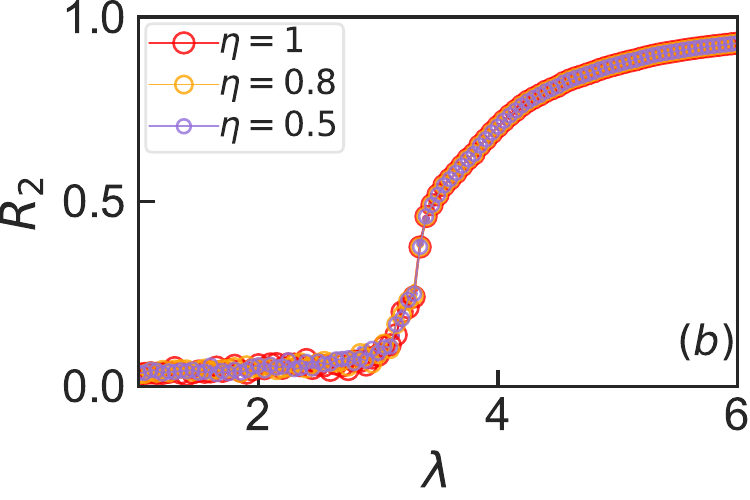}\\
	\end{tabular}{}
	\vspace{-0.5cm}
	\caption{(Color online) (a) $R_1{-}\lambda$ and (b) $R_2{-}\lambda$ profiles for an all-to-all dyadic network (of 1-simplices) simulated with Gaussian $g(\omega)$ with $\Delta{=}1$, $N{=}10^3$, $\alpha{=1}$ and $\kappa=1$.}
	\label{figureS1}
\end{figure}

 Fig.~\ref{figureS1} illustrates that both $R_1$ and $R_2$ desynchronize espousing a second-order transition route. Nonetheless, the maximum value to which $R_1$ plateaus in the synchronous state is characterized by the asymmetry probability $\eta$. Moreover, $R_1$ does not see synchronization for the forward continuation of $\lambda$.

\subsection{Robust desynchronization with unimodal frequency distributions $g(\omega)$}
\begin{figure}[ht!]
	\centering
	\begin{tabular}{cc}
	\includegraphics[height=3.2cm,width=4cm]{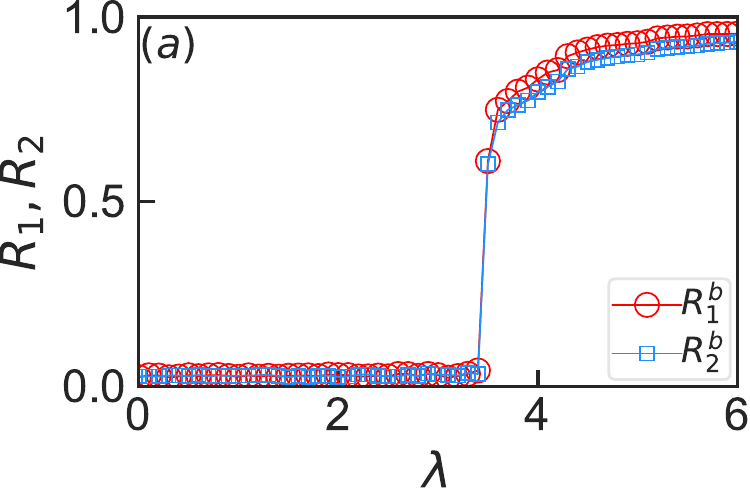}&
	\includegraphics[height=3.2cm,width=4cm]{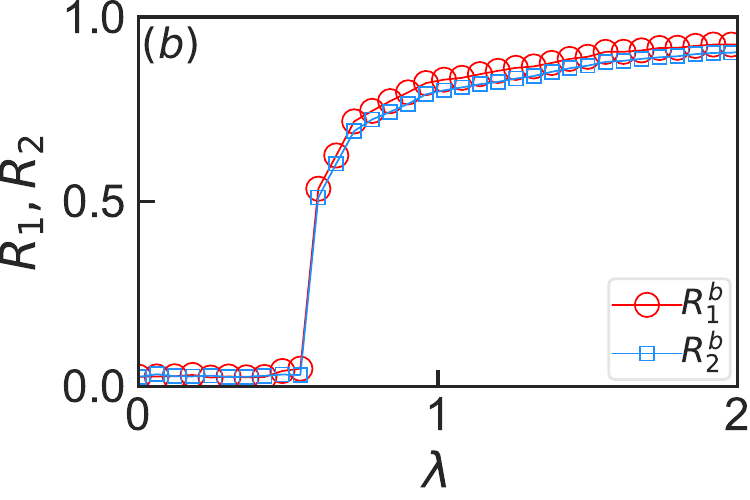}\\
	\end{tabular}{}
	\vspace{-0.5cm}
	\caption{(Color online) $R_1{-}\lambda$ and $R_2{-}\lambda$ profiles for a random 2-simplex network simulated with (a) Gaussian $g(\omega)$ with $\Delta{=}1$ and (b) Lorentzian $g(\omega)$ with $\Delta{=}0.1$.\ Other network parameters are $N{=}10^3$, $\langle d^{[2]}\rangle{=8}$, $\varepsilon{=1}$, $\mu=1$ and $\eta=1$.}
	\label{figureS2}
\end{figure}

\begin{figure}[ht!]
	\centering
	\begin{tabular}{cc}
	\includegraphics[height=3.2cm,width=4cm]{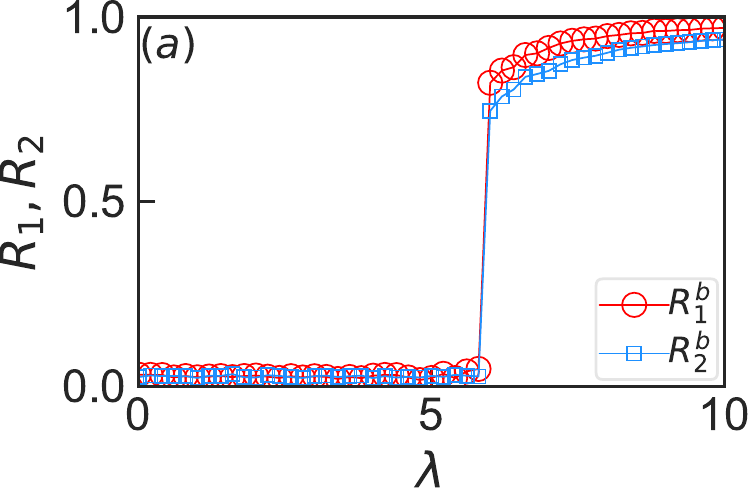}&
	\includegraphics[height=3.2cm,width=4cm]{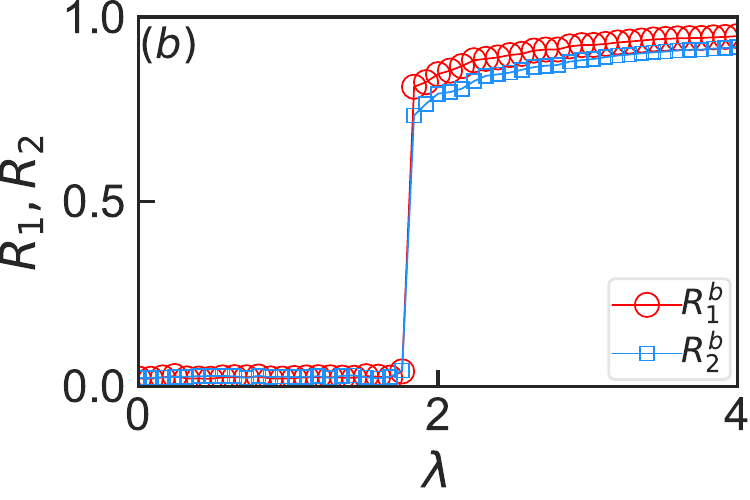}\\
	\end{tabular}{}
	\vspace{-0.5cm}
	\caption{(Color online) $R_1{-}\lambda$ and $R_2{-}\lambda$ profiles for a random 3-simplex network simulated with (a) Gaussian $g(\omega)$ with $\Delta{=}1$ and (b) Lorentzian $g(\omega)$ with $\Delta{=}0.1$.\ Other network parameters are $N{=}10^3$, $\langle d^{[3]}\rangle{=12}$, $\sigma{=1}$, $\nu=1$ and $\eta=1$.}
	\label{figureS3}
\end{figure}

A manifestation of the first-order desynchronization in a random pure 2-simplex and 3-simplex networks corresponding to a Gaussian $g(\omega){=}\frac{1}{\Delta\sqrt{2\pi}}\exp(-\omega^2/2\Delta^2)$ and a Lorentzian $g(\omega){=}\frac{\Delta}{\pi(\omega^2+\Delta^2)}$, where $\Delta=$ half width at half maximum, are shown in Fig.~\ref{figureS2} and Fig.~\ref{figureS3}, respectively. The abrupt desynchronization takes place at different critical coupling strength for the two symmetric distributions and forward transition to synchronization does not occur for any $\lambda>0$.

\subsection{\textbf{Order parameters for the adaptive 2-simplicial complex}}
The expressions of the order parameters for the 2-simplicial complex treated for the uniform distribution $g(\omega){=}\frac{1}{2\Delta}$:
\begin{align}\label{predict}
	\begin{split}
	&R_1 = 2(2\eta-1)\frac{bq}{\Delta} \bigg[ \frac{1}{3} \sin\bigg\lgroup\frac{3}{4} \arcsin\bigg\lgroup\frac{\Delta}{bq}\bigg\rgroup\bigg\rgroup + \frac{1}{5} \sin\bigg\lgroup\frac{5}{4} \arcsin\bigg\lgroup\frac{\Delta}{bq}\bigg\rgroup\bigg\rgroup \bigg],\nonumber
	\end{split}\\
	\begin{split}
	&R_2 = \frac{\sqrt{2}b^2q^2}{3\Delta^2}\Bigg\lgroup1-\sqrt{1-\frac{\Delta^2}{b^2q^2}}\Bigg\rgroup\Bigg\lgroup2+ \sqrt{1-\frac{\Delta^2}{b^2q^2}}\Bigg\rgroup \sqrt{1+\sqrt{1-\frac{\Delta^2}{b^2q^2}}},
	\end{split}
\end{align}
where $b={\varepsilon}/{2\mu}$ and $q=\lambda R_2^2$.

\subsection{Order parameters for the adaptive 3-simplicial complex}
The fixed points of the locked oscillators related to the Hebbian learning ruled 3-simplicial complex are given by $\sin(6\theta_i^*){=}\omega_i/d\lambda R_2^3;\ d{=}\sigma/(2\nu)$,
which can be further simplified to $4\cos^3(2\theta_i^*) - 3\cos(2\theta_i^*){=}\sqrt{1-\omega^2_i/u^2}$, where $u=d\lambda R_2^3$. This equation that is cubic in $\cos(2\theta^*_i)$ supplies us with following three roots
\begin{align}\label{3fixed_points}
	&\cos(2\theta^*_1)= \frac{1}{2}\big[\sqrt[3]{x+iy} + \sqrt[-3]{x+iy}\big],\nonumber \\
	&\cos(2\theta^*_{2,3})= \frac{1}{4}\big[{\pm(i\sqrt{3}\mp 1) \sqrt[3]{x+iy} \mp (i\sqrt{3}\pm1}){\sqrt[-3]{x+iy}}\big],
\end{align}
where $x{=}\sqrt{1-{\omega^2_i}/{u^2}},\ y{=}{\omega_i}/{u}$. Only the fixed point given by $\cos(2\theta^*_1)$ offers a viable solution for the order parameters. Hence, the order parameters for the locked oscillators in continuum limit can be expressed as
\begin{eqnarray}\label{opara_m}
R_m = \int_{|\omega|\leq u}\mathrm{d\omega}g(\omega)\int_{0}^{2\pi}\mathrm{d\theta} e^{\imath m \theta} [\eta\delta(\theta-\theta^\star_1) + (1-\eta) \delta(\theta-\theta^\star_1-\pi)].
\end{eqnarray}
Further working out for $m=1,2$, we obtain the real order parameters as
\begin{eqnarray}\label{opara_real}
	&R_1 = \frac{(2\eta-1)}{\sqrt2}\int\displaylimits_{|\omega|\leq u}\mathrm{d\omega}g(\omega) \sqrt{1+\cos(2\theta^*_1)},\nonumber \\
	&R_2 = \int\displaylimits_{|\omega|\leq u}\mathrm{d\omega}g(\omega) \cos(2\theta^*_1).
\end{eqnarray}
Considering all the oscillators locked for a uniform distribution $g(\omega){=}\frac{1}{2\Delta}$ and substituting $\cos(2\theta^*_1)$ from Eq.(\ref{3fixed_points}), we obtain
\begin{eqnarray}\label{3simpR}
	&R_1 =\frac{(2\eta-1)}{2\sqrt2 \Delta}\int_{-\Delta}^{\Delta} \mathrm{d\omega} \sqrt{1+\frac{1}{2}\big[\sqrt[3]{x+iy} + \sqrt[-3]{x+iy}\big]},\nonumber\\
	&R_2=\frac{1}{4\Delta}\int_{-\Delta}^{\Delta} \mathrm{d\omega} \big[\sqrt[3]{x+iy} + \sqrt[-3]{x+iy}\big], 
\end{eqnarray}
where $x{=}\sqrt{1-{\omega^2}/{u^2}},\ y{=}{\omega}/{u}$. Eq.(\ref{3simpR}) can further mathematically be simplified to
\begin{align}
	&R_1 = \frac{3u(2\eta-1)}{35\sqrt2\Delta} \sec\bigg\lgroup\frac{l}{6}\bigg\rgroup \bigg[7\sin\bigg\lgroup\frac{5l}{6}\bigg\rgroup + 5\sin\bigg\lgroup\frac{7l}{6}\bigg\rgroup\bigg] \sqrt{1+\cos\bigg\lgroup\frac{l}{3}\bigg\rgroup}, \nonumber\\
	&R_2 = \frac{3u}{\Delta} \sin\bigg\lgroup\frac{1}{3}\arcsin{\frac{\Delta}{u}}\bigg\rgroup \cos^3\bigg\lgroup\frac{1}{3}\arcsin{\frac{\Delta}{u}}\bigg\rgroup,
\end{align}
where $l = \arcsin({\Delta}/{u})$ and $u=d\lambda R_2^3$.